%% file: Douplii-Guo_PolEncoding-arxiv1.tex
\documentclass[aps,prd,onecolumn,12pt,longbibliography]{revtex4-2}
\usepackage{amsmath,amsthm}
\usepackage{amsfonts}
\usepackage{amssymb}
\usepackage{amsbsy,bm}

\theoremstyle{plain}

\theoremstyle{definition}

\theoremstyle{remark}

\numberwithin{equation}{section}

\renewcommand{\mathit}{\bm}
\date{July 7, 2025}
\usepackage{natbib}

\begin{document}

\title{Polyadic encryption}

\author{Steven Duplij}
\email{douplii@uni-muenster.de, http://www.uni-muenster.de/IT.StepanDouplii}

\affiliation{Center for Information Technology,
University of M\"unster,
48149 M\"unster,
Germany}
\affiliation{
Yantai Research Institute,
Harbin Engineering University,
265615 Yantai, China}

\author{Qiang Guo}
\email{guoqiang292004@163.com, guoqiang@hrbeu.edu.cn}

\affiliation{
College of Information and Communication Engineering,
Harbin Engineering University,
150001 Harbin, China}

\begin{abstract}
\input{Douplii-Guo_PolEncoding-abs}

\end{abstract}

\maketitle

\input{Douplii-Guo_PolEncoding-sw1}
\newpage

\input{Douplii-Guo_PolEncoding1.bbl}
\end{document}

%% file: Douplii-Guo_PolEncoding-abs.tex

\noindent A novel original procedure of encryption/decryption based on the
polyadic algebraic structures and on signal processing methods is proposed.
First, we use signals with integer amplitudes to send information. Then we use
polyadic techniques to transfer the plaintext into series of special integers.
The receiver restores the plaintext using special rules and systems of equations.

\smallskip


%% file: Douplii-Guo_PolEncoding-sw1.tex
\section{Introduction}

We propose a new approach to transfer hidden information in (continuous-time
discrete-valued) signal processing (see, e.g. \cite{oppenheim,wang}) by
considering the parameters of signals not as the ordinary integers
\cite{berber,ped-ull}, but as a special kind of integer numbers, polyadic
integers, introduced in \cite{duplij2022}. The polyadic integers form
polyadic $\left(  m,n\right)  $-rings having (or closed with respect to) $m$
additions and $n$ multiplications \cite{lee/but}. In this way, preservation of
the property to be in the same polyadic ring after signal processing will give
various restrictions on the signal parameters. The main idea is to use
these restrictions (as equations) to encrypt and decrypt a series of ordinary
numbers using sets of signals prepared in special ways.

\section{Polyadic rings}

We first remind that the polyadic ring or $\left(  m,n\right)  $-ring is a set
with $m$-ary addition (being a $m$-ary group) and $n$-ary multiplication
(being a $n$-ary semigroup), which are connected by the polyadic distributive
law \cite{lee/but}. A polyadic ring is nonderived, if its full operations
cannot be obtained as repetition of binary operations. As the simplest example
of the binary ring or $\left(  2,2\right)  $-ring is the set ordinary integers
$\mathbb{Z}$, the example of $\left(  m,n\right)  $-ring is the set of
polyadic integers $\mathbb{Z}_{\left(  m,n\right)  }$ \cite{duplij2022}.
The concrete realization of polyadic integers is the set of representative of
the conguence class (residue class) of an integer $a$ modulo $b$ (with both
$a$ and $b$ fixed)%
\begin{equation}
\mathbb{Z}_{\left(  m,n\right)  }^{\left[  a,b\right]  }=\left[  \left[
a\right]  \right]  _{b}=\left\{  \left\{  a+b\cdot k\right\}  \mid
k\in\mathbb{Z},\ \ a\in\mathbb{Z}_{+},\ b\in\mathbb{N},\ 0\leq a\leq
b-1\right\}  . \label{ab}%
\end{equation}
We denote a representative element by $x_{k}=x_{k}^{\left[  a,b\right]
}=a+b\cdot k$, which are polyadic intergers $\left\{  x_{k}^{\left[
a,b\right]  }\right\}  \in\mathbb{Z}_{\left(  m,n\right)  }^{\left[
a,b\right]  }$, because only $m$ additions and $n$ multiplications are
possible in the nonderived case%
\begin{align}
\overset{m}{\overbrace{x_{k_{1}}+x_{k_{2}}+\ldots+x_{k_{m}}}}  &
\in\mathbb{G}_{add\left(  m\right)  }^{\left[  a,b\right]  }\subset
\mathbb{Z}_{\left(  m,n\right)  }^{\left[  a,b\right]  },\label{nu}\\
\overset{n}{\overbrace{x_{k_{1}}x_{k_{2}}\ldots x_{k_{n}}}}  &  \in
\mathbb{S}_{mult\left(  n\right)  }^{\left[  a,b\right]  }\subset
\mathbb{Z}_{\left(  m,n\right)  }^{\left[  a,b\right]  },\ \ k_{i}%
\in\mathbb{Z}, \label{mu}%
\end{align}
where $\mathbb{G}_{add\left(  m\right)  }^{\left[  a,b\right]  }\mathbb{\ }%
$and $\mathbb{S}_{mult\left(  n\right)  }^{\left[  a,b\right]  }$ are the
$m$-ary additive group and $n$-ary multiplicative semigroup of the polyadic
ring $\mathbb{Z}_{\left(  m,n\right)  }^{\left[  a,b\right]  }$.

It follows from (\ref{nu})--(\ref{mu}), that more generally the $m$-admissible
sum consists of $\ell_{m}\left(  m-1\right)  +1$ summands and the
$n$-admissible product contains $\ell_{n}\left(  n-1\right)  +1$ elements,
where $\ell_{m}$ is a number of $m$-ary additions ($m$-polyadic power) and
$\ell_{n}$ is a number of $n$-ary multiplications ($n$-polyadic power).
Therefore, in general%
\begin{align}
\overset{\ell_{m}\left(  m-1\right)  +1}{\overbrace{x_{k_{1}}+x_{k_{2}}%
+\ldots+x_{k_{m}}}}  &  \in\mathbb{G}_{add\left(  m\right)  }^{\left[
a,b\right]  }\subset\mathbb{Z}_{\left(  m,n\right)  }^{\left[  a,b\right]
},\label{nu1}\\
\overset{\ell_{n}\left(  n-1\right)  +1}{\overbrace{x_{k_{1}}x_{k_{2}}\ldots
x_{k_{n}}}}  &  \in\mathbb{S}_{mult\left(  n\right)  }^{\left[  a,b\right]
}\subset\mathbb{Z}_{\left(  m,n\right)  }^{\left[  a,b\right]  },\ \ k_{i}%
\in\mathbb{Z}. \label{mu1}%
\end{align}
$\ell_{m}\left(  m-1\right)  +1$

For instance, in the residue (congruence) class%
\begin{equation}
\left[  \left[  3\right]  \right]  _{4}=\left\{  \ldots
-25,-21,-17,-13,-9,-5,-1,3,7,11,15,19,23,27,31,35,39\ldots\right\}  \label{34}%
\end{equation}
we can add $4\ell_{m}+1$ representatives and multiply $2\ell_{n}+1$
representatives ($\ell_{m},\ell_{n}$ are polyadic powers) to retain in the
same class $\left[  \left[  3\right]  \right]  _{4}$. If, for example, we take
$\ell_{m}=2$, $\ell_{n}=3$, then we obtain closeness of polyadic operations%
\begin{align}
\left(  7+11+15+19+23\right)  -5-9-13-1  &  =\allowbreak47=3+4\cdot
11\in\left[  \left[  3\right]  \right]  _{4},\\
\left(  \left(  7\cdot3\cdot11\right)  \cdot19\cdot15\right)  \cdot31\cdot27
&  =\allowbreak55\,103\,895=3+4\cdot13\,775\,973\in\left[  \left[  3\right]
\right]  _{4}.
\end{align}

Thus, we cannot add and multiply arbitrary quantities of representatives in
$\left[  \left[  3\right]  \right]  _{4}$, only the admissible ones. This
means that $\left[  \left[  3\right]  \right]  _{4}$ is really the polyadic
$\left(  5,3\right)  $-ring $\mathbb{Z}_{\left(  5,3\right)  }^{\left[
3,4\right]  }$.

In general, a congruence class $\left[  \left[  a\right]  \right]  _{b}$ is a
polyadic ring $\mathbb{Z}_{\left(  m,n\right)  }^{\left[  a,b\right]  }$, if
the following relations hold valid \cite{duplij2022}%
\begin{equation}
\left(  m-1\right)  \dfrac{a}{b}=I^{\left(  m\right)  }\left(  a,b\right)
=I=\operatorname{integer}, \label{am}%
\end{equation}%
\begin{equation}
\dfrac{a^{n}-a}{b}=J^{\left(  n\right)  }\left(  a,b\right)
=J=\operatorname{integer}, \label{an}%
\end{equation}
where $I,J$ are called a (polyadic) shape invariants of the congruence class
$\left[  \left[  a\right]  \right]  _{b}$, e.g., for the congruence class
$\left[  \left[  3\right]  \right]  _{4}$ the shape invariants (\ref{am}%
)--(\ref{an}) are $I=3$ and $J=6$, correspondingly.

In TABLE \ref{T1} the mapping%
\begin{equation}
\Phi_{\left(  m,n\right)  }^{\left[  a,b\right]  }:\left(  a,b\right)
\longrightarrow\left(  m,n\right)  \label{f}%
\end{equation}
of the congruence class parameters to the polyadic ring arities (we call it
the arity shape) and shape invariants is presented for their lowest values.
The arity shape mapping (\ref{f}) is injective and non-surjective (empty
cells), and it cannot be expressed in closed form. Moreover,
e.g., the congruence classes $\left[  \left[  2\right]  \right]  _{4}$,
$\left[  \left[  2\right]  \right]  _{8}$, $\left[  \left[  3\right]  \right]
_{9}$, $\left[  \left[  4\right]  \right]  _{8}$, $\left[  \left[  6\right]
\right]  _{8}$ and $\left[  \left[  6\right]  \right]  _{9}$ do not correspond
to any ring, while the same $\left(  6,5\right)  $-ring can be described by
different congruence classes $\left[  \left[  2\right]  \right]  _{5}$,
$\left[  \left[  3\right]  \right]  _{5}$, $\left[  \left[  2\right]  \right]
_{10}$, and $\left[  \left[  8\right]  \right]  _{10}$.

The polyadic arity shape $\Phi_{\left(  m,n\right)  }^{\left[  a,b\right]  }$
(\ref{f}) is the main tool in the encryption/decryption procedure, described below.

\begin{table}[h]
\caption{The polyadic ring $\mathbb{Z}_{\left(  m,n\right)  }^{\left[
a,b\right]  }$ of the fixed residue class $\left[  \left[  a\right]  \right]
_{b}$: the arity shape $\Phi_{\left(  m,n\right)  }^{\left[  a,b\right]  }$. }%
\label{T1}
\begin{center}
{\tiny
\begin{tabular}
[c]{||c||c|c|c|c|c|c|c|c|c||}\hline\hline
$a\setminus b$ & 2 & 3 & 4 & 5 & 6 & 7 & 8 & 9 & 10\\\hline\hline
1 & $%
\begin{array}
[c]{c}%
m=\mathbf{3}\\
n=\mathbf{2}\\
I=1\\
J=0
\end{array}
$ & $%
\begin{array}
[c]{c}%
m=\mathbf{4}\\
n=\mathbf{2}\\
I=1\\
J=0
\end{array}
$ & $%
\begin{array}
[c]{c}%
m=\mathbf{5}\\
n=\mathbf{2}\\
I=1\\
J=0
\end{array}
$ & $%
\begin{array}
[c]{c}%
m=\mathbf{6}\\
n=\mathbf{2}\\
I=1\\
J=0
\end{array}
$ & $%
\begin{array}
[c]{c}%
m=\mathbf{7}\\
n=\mathbf{2}\\
I=1\\
J=0
\end{array}
$ & $%
\begin{array}
[c]{c}%
m=\mathbf{8}\\
n=\mathbf{2}\\
I=1\\
J=0
\end{array}
$ & $%
\begin{array}
[c]{c}%
m=\mathbf{9}\\
n=\mathbf{2}\\
I=1\\
J=0
\end{array}
$ & $%
\begin{array}
[c]{c}%
m=\mathbf{10}\\
n=\mathbf{2}\\
I=1\\
J=0
\end{array}
$ & $%
\begin{array}
[c]{c}%
m=\mathbf{11}\\
n=\mathbf{2}\\
I=1\\
J=0
\end{array}
$\\\hline
2 &  & $%
\begin{array}
[c]{c}%
m=\mathbf{4}\\
n=\mathbf{3}\\
I=2\\
J=2
\end{array}
$ &  & $%
\begin{array}
[c]{c}%
m=\mathbf{6}\\
n=\mathbf{5}\\
I=2\\
J=6
\end{array}
$ & $%
\begin{array}
[c]{c}%
m=\mathbf{4}\\
n=\mathbf{3}\\
I=1\\
J=1
\end{array}
$ & $%
\begin{array}
[c]{c}%
m=\mathbf{8}\\
n=\mathbf{4}\\
I=2\\
J=2
\end{array}
$ &  & $%
\begin{array}
[c]{c}%
m=\mathbf{10}\\
n=\mathbf{7}\\
I=2\\
J=14
\end{array}
$ & $%
\begin{array}
[c]{c}%
m=\mathbf{6}\\
n=\mathbf{5}\\
I=1\\
J=3
\end{array}
$\\\hline
3 &  &  & $%
\begin{array}
[c]{c}%
m=\mathbf{5}\\
n=\mathbf{3}\\
I=3\\
J=6
\end{array}
$ & $%
\begin{array}
[c]{c}%
m=\mathbf{6}\\
n=\mathbf{5}\\
I=3\\
J=48
\end{array}
$ & $%
\begin{array}
[c]{c}%
m=\mathbf{3}\\
n=\mathbf{2}\\
I=1\\
J=1
\end{array}
$ & $%
\begin{array}
[c]{c}%
m=\mathbf{8}\\
n=\mathbf{7}\\
I=3\\
J=312
\end{array}
$ & $%
\begin{array}
[c]{c}%
m=\mathbf{9}\\
n=\mathbf{3}\\
I=3\\
J=3
\end{array}
$ &  & $%
\begin{array}
[c]{c}%
m=\mathbf{11}\\
n=\mathbf{5}\\
I=3\\
J=24
\end{array}
$\\\hline
4 &  &  &  & $%
\begin{array}
[c]{c}%
m=\mathbf{6}\\
n=\mathbf{3}\\
I=4\\
J=12
\end{array}
$ & $%
\begin{array}
[c]{c}%
m=\mathbf{4}\\
n=\mathbf{2}\\
I=2\\
J=2
\end{array}
$ & $%
\begin{array}
[c]{c}%
m=\mathbf{8}\\
n=\mathbf{4}\\
I=4\\
J=36
\end{array}
$ &  & $%
\begin{array}
[c]{c}%
m=\mathbf{10}\\
n=\mathbf{4}\\
I=4\\
J=28
\end{array}
$ & $%
\begin{array}
[c]{c}%
m=\mathbf{6}\\
n=\mathbf{3}\\
I=2\\
J=6
\end{array}
$\\\hline
5 &  &  &  &  & $%
\begin{array}
[c]{c}%
m=\mathbf{7}\\
n=\mathbf{3}\\
I=5\\
J=20
\end{array}
$ & $%
\begin{array}
[c]{c}%
m=\mathbf{8}\\
n=\mathbf{7}\\
I=5\\
J=11160
\end{array}
$ & $%
\begin{array}
[c]{c}%
m=\mathbf{9}\\
n=\mathbf{3}\\
I=5\\
J=15
\end{array}
$ & $%
\begin{array}
[c]{c}%
m=\mathbf{10}\\
n=\mathbf{7}\\
I=5\\
J=8680
\end{array}
$ & $%
\begin{array}
[c]{c}%
m=\mathbf{3}\\
n=\mathbf{2}\\
I=1\\
J=2
\end{array}
$\\\hline
6 &  &  &  &  &  & $%
\begin{array}
[c]{c}%
m=\mathbf{8}\\
n=\mathbf{3}\\
I=6\\
J=30
\end{array}
$ &  &  & $%
\begin{array}
[c]{c}%
m=\mathbf{6}\\
n=\mathbf{2}\\
I=3\\
J=3
\end{array}
$\\\hline
7 &  &  &  &  &  &  & $%
\begin{array}
[c]{c}%
m=\mathbf{9}\\
n=\mathbf{3}\\
I=7\\
J=42
\end{array}
$ & $%
\begin{array}
[c]{c}%
m=\mathbf{10}\\
n=\mathbf{4}\\
I=7\\
J=266
\end{array}
$ & $%
\begin{array}
[c]{c}%
m=\mathbf{11}\\
n=\mathbf{5}\\
I=7\\
J=1680
\end{array}
$\\\hline
8 &  &  &  &  &  &  &  & $%
\begin{array}
[c]{c}%
m=\mathbf{10}\\
n=\mathbf{3}\\
I=8\\
J=56
\end{array}
$ & $%
\begin{array}
[c]{c}%
m=\mathbf{6}\\
n=\mathbf{5}\\
I=4\\
J=3276
\end{array}
$\\\hline
9 &  &  &  &  &  &  &  &  & $%
\begin{array}
[c]{c}%
m=\mathbf{11}\\
n=\mathbf{3}\\
I=9\\
J=72
\end{array}
$\\\hline\hline
\end{tabular}
}
\end{center}
\end{table}

\section{Polyadic encryption/decryption procedure}

Let us consider the initial plaintext as a series of ordinary integer numbers
(any plaintext can be transformed to that by the corresponding encoding
procedure)%
\begin{equation}
\mathbf{T}=y_{1},y_{2},\ldots y_{r},\ \ \ \ \ \ y_{j}\in\mathbb{Z}. \label{t}%
\end{equation}

We propose a general encryption/decryption procedure, when each of $y=y_{j}$
is connected with the various parameters of signal series, and the latter are
transfered to the receiver, who then restores $y$ using special rules and
systems of equations known to him only.

The main idea is to examine such signals which have parameters as polyadics
integers, that is they are in the polyadic ring $\mathbb{Z}_{\left(
m,n\right)  }^{\left[  a,b\right]  }$ (\ref{ab}). This can be treated as a
polyadic generalization of the (binary) discretization technique (in which the
parameters are ordinary integers $\mathbb{Z}$), and so we call it the polyadic
discretization. Its crucial new feature is the possibility to transfer
information (e.g. arities, minimal allowed number of additions and
multiplications) using signal parameters, as it will be shown below.

Here, we apply this idea to signal amplitudes (such signals are called the
continuous-time discrete-valued or quantized analog signal
\cite{mus/lis,kar/ash/nij}) and their addition only. This means that we look
on the additive part of the polyadic ring $\mathbb{Z}_{\left(  m,n\right)
}^{\left[  a,b\right]  }$ which is a nonderived (allowed to add exactly $m$ terms, no fewer) $m$-ary group $\mathbb{G}_{add\left(  m\right)  }^{\left[  a,b\right]
}$ (\ref{nu}). In this way, we denote the single $i$th signal shape as%
\begin{equation}
\Psi_{i}^{\left(  \ell_{f}\right)  }=A_{i}^{\left(  \ell_{f}\right)  }\cdot
f^{\left(  \ell_{f}\right)  }\left(  t\right)  , \label{fa}%
\end{equation}
where $A_{i}^{\left(  \ell_{f}\right)  }$ is the amplitude of the normalized
(in some manner) $i$th signal $f^{\left(  \ell_{f}\right)  }\left(  t\right)
$, $t$ is time, and the natural $\ell_{f}\in\mathbb{N}$ corresponds to the
special kind of signal (by consequent numerating sine/cosine, triangular,
rectangular, etc.).

First, we assume that the amplitude $A_{i}^{\left(  \ell_{f}\right)  }$ is in
polyadic ring, i.e., it is a representative of the congruence class $\left[
\left[  a\right]  \right]  _{b}$, $\ b\in\mathbb{N},\ 0\leq a\leq b-1$, and
therefore, it has the form (\ref{ab})%
\begin{equation}
A_{i}^{\left(  \ell_{f}\right)  }=a+b\cdot k_{i}^{\left(  \ell_{f}\right)
}\in\mathbb{G}_{add\left(  m\right)  }^{\left[  a,b\right]  }\subset
\mathbb{Z}_{\left(  m,n\right)  }^{\left[  a,b\right]  },\ \ \ \ \ \ \ k_{i}%
^{\left(  \ell_{f}\right)  }\in\mathbb{Z}. \label{al}%
\end{equation}

Second, we identify the number of signal species $\ell_{f}$ with the
$m$-polyadic power $\ell_{m}$ from (\ref{nu1})%
\begin{equation}
\ell_{f}=\ell_{m}. \label{ll}%
\end{equation}

In this picture, for the signal species $\ell_{f}$ we prepare the sum of
$\ell_{f}\left(  m-1\right)  +1$ signals as%
\begin{equation}
\Psi_{tot}^{\left(  \ell_{f}\right)  }=\sum_{i=1}^{\ell_{f}\left(  m-1\right)
+1}\Psi_{i}^{\left(  \ell_{f}\right)  }=A_{tot}^{\left(  \ell_{f}\right)
}\cdot f^{\left(  \ell_{f}\right)  }\left(  t\right)  \label{af}%
\end{equation}
where the total amplitude $A_{tot}^{\left(  \ell_{f}\right)  }$ becomes
different for distinct species $\ell_{f}$ and after usage of (\ref{al}) has
the general form%
\begin{align}
A_{tot}^{\left(  \ell_{f}\right)  }  &  =\sum_{i=1}^{\ell_{f}\left(
m-1\right)  +1}A_{i}^{\left(  \ell_{f}\right)  }=a\cdot\left(  \ell_{f}\left(
m-1\right)  +1\right)  +b\cdot K\left(  m,\ell_{f}\right)  \in\mathbb{G}%
_{add\left(  m\right)  }^{\left[  a,b\right]  }\subset\mathbb{Z}_{\left(
m,n\right)  }^{\left[  a,b\right]  },\label{at}\\
K\left(  m,\ell_{f}\right)   &  =\sum_{i=1}^{\ell_{f}\left(  m-1\right)
+1}k_{i}^{\left(  \ell_{f}\right)  }. \label{k}%
\end{align}

Thus, we observe that the total amplitude (\ref{at}) of the signal (\ref{fa})
contains all parameters of the $m$-ary group $\mathbb{G}_{add\left(  m\right)
}^{\left[  a,b\right]  }$ (the additive part of the polyadic ring
$\mathbb{Z}_{\left(  m,n\right)  }^{\left[  a,b\right]  }$ (\ref{ab})). This
allows us to use the combination of signals (in the above particular case
sums) to transfer securely the plaintext variables $y_{j}$ (\ref{t}) from
sender to recepient, if we encode each of them $y=y_{j}$ by the polyadic ring
parameters $y\longrightarrow\left(  a,b,m\right)  $. The recepient obtains the
set of the total amplitudes $A_{tot}^{\left(  \ell_{f}\right)  }$ and treats
(\ref{at}) as the system of equations for parameters $\left(  a,b,m\right)  $,
and then after the decoding $\left(  a,b,m\right)  \longrightarrow
y\longrightarrow y=y_{j}$ for each $j$ obtains the initial plaintext
$\mathbf{T}$ (\ref{t}). Schematically, we can present the proposed
encryption/decryption procedure as%
\begin{align}
&  \text{\textsf{snd}:\ }y\overset{\text{encoding by snd}}{\longrightarrow
}\left(  a,b,m\right)  \overset{\text{summing}}{\longrightarrow}%
A_{tot}^{\left(  \ell_{f}\right)  }\overset{\text{preparing signals}%
}{\longrightarrow}\Psi_{tot}^{\left(  \ell_{f}\right)  }\overset
{\text{transferring to rcp}}{\longrightarrow}\\
&  \text{\textsf{rcp}: system of equations }A_{tot}^{\left(  \ell_{f}\right)
}\text{ (\ref{at})}\overset{\text{solving by rcp}}{\longrightarrow}\left(
a,b,m\right)  \overset{\text{decoding by rcp}}{\longrightarrow}y
\end{align}

The security of this procedure is goverened not by one key, as in the standard
cases, but by the system (\ref{at}) and the TABLE \ref{T1}, and by
connection between a kind of signal $\ell_{f}$ and $m$-polyadic
power (\ref{ll}), which are all unknown to the third party.

\section{Example}

Let us consider a concrete example of the proposed encryption/decryption
procedure for the congruence class $\left[  \left[  a\right]  \right]  _{b}$
and one kind of signal $\ell_{f}=\ell_{m}\equiv\ell$, where $\ell$ is
$m$-polyadic power. Each of such class gives (by TABLE \ref{T1}) the arity $m$ as
the plaintext entry to transfer $y=m$. Next we should choose the shape of the
function $K\left(  m,\ell\right)  $ (\ref{k}), which is, in general,
arbitrary. In the simplest case, we take the same linear function%
\begin{equation}
k_{i}^{\left(  \ell\right)  }=i-1, \label{i}%
\end{equation}
for all $m$-polyadic powers $\ell$, but any functional dependence in
(\ref{i}) can be chosen, and it is different for different $\ell$, which increases
security of the procedure. The choice (\ref{i}) gives%
\begin{equation}
K\left(  m,\ell\right)  =\frac{1}{2}\ell\left(  m-1\right)  \left(
\ell\left(  m-1\right)  +1\right)  .
\end{equation}

So the total amplitudes for different polyadic powers become%
\begin{equation}
A_{tot}^{\left(  \ell\right)  }\equiv B_{\ell}=a\cdot\left(  \ell\left(
m-1\right)  +1\right)  +\frac{b}{2}\ell\left(  m-1\right)  \left(  \ell\left(
m-1\right)  +1\right)  \cdot\label{bl}%
\end{equation}

The recepient obtains the set of signals with amplitudes (\ref{bl}) as
polyadic integers%
\begin{equation}
\Psi_{tot}^{\left(  \ell\right)  }=B_{\ell}\cdot f^{\left(  \ell\right)
}\left(  t\right)  . \label{bf}%
\end{equation}

To obtain the values of three variables $\left(  a,b,m\right)  $, one needs three
equations, i.e., three total amplitudes with different arbitrary polyadic powers
$\ell=\ell_{1},\ell_{2},\ell_{3}\in\mathbb{N}$. Because the general solution
is too cumbersome, we choose the first three consequent polyadic powers
$\ell=1,2,3$, while any three natural numbers are possible to increase the
security. This gives the following system of quadratic equations%
\begin{align}
a\cdot m+b\cdot\frac{\left(  m-1\right)  m}{2}  &  =B_{1},\label{ab1}\\
a\cdot\left(  2m-1\right)  +b\cdot\left(  m-1\right)  \left(  2m-1\right)   &
=B_{2},\label{ab2}\\
a\cdot\left(  3m-2\right)  +b\cdot\frac{3\left(  m-1\right)  \left(
3m-2\right)  }{2}  &  =B_{3}. \label{ab3}%
\end{align}

The general solution of the system is%
\begin{equation}
m=\frac{7B_{1}-4B_{2}+B_{3}\mp\sqrt{B_{1}^{2}-8B_{2}B_{1}-2B_{3}B_{1}%
+16B_{2}^{2}+B_{3}^{2}-8B_{2}B_{3}}}{4\left(  3B_{1}-3B_{2}+B_{3}\right)  },
\label{mb}%
\end{equation}%
\begin{equation}
a=3B_{1}-3B_{2}+B_{3}, \label{ma}%
\end{equation}%
\begin{align}
b  &  =\frac{1}{\left(  3B_{1}-3B_{2}+B_{3}\right)  \left(  2B_{2}-B_{1}%
-B_{3}\right)  }\times\nonumber\\
&  \left[  \ \left(  11B_{1}^{2}-16B_{2}B_{1}+6B_{3}B_{1}-4B_{2}^{2}%
+4B_{2}B_{3}-B_{3}^{2}\right)  \left(  3B_{1}-3B_{2}+B_{3}\right)  \right.
\nonumber\\
&  -\frac{105}{2}B_{1}^{3}+\frac{333}{2}B_{1}^{2}B_{2}-162B_{1}B_{2}%
^{2}+48B_{2}^{3}-\frac{113}{2}B_{1}^{2}B_{3}\nonumber\\
&  +107B_{1}B_{2}B_{3}-46B_{2}^{2}B_{3}-\frac{35}{2}B_{1}B_{3}^{2}+\frac
{29}{2}B_{2}B_{3}^{2}-\frac{3}{2}B_{3}^{3}\nonumber\\
&  \pm\left(  \frac{15}{2}B_{1}^{2}-\frac{39}{2}B_{1}B_{2}+12B_{2}^{2}%
+7B_{1}B_{3}-\frac{17}{2}B_{2}B_{3}+\frac{3}{2}B_{3}^{2}\right)
\times\nonumber\\
&  \left.  \sqrt{B_{1}^{2}-8B_{2}B_{1}-2B_{3}B_{1}+16B_{2}^{2}+B_{3}%
^{2}-8B_{2}B_{3}}\ \right]  . \label{bb}%
\end{align}

The sign in (\ref{mb}) and (\ref{bb}) should be chosen so that the solutions
are ordinary integers.

In this particular case, for instance, the congruence class $\left[  \left[
3\right]  \right]  _{4}$ (\ref{34}), has $5$-ary addition, as follows from
TABLE \ref{T1}. So if the sender wants to securely submit one element $y=m=5$
from his plainintext (\ref{t}), he applies the proposed encryption procedure
and prepares three sums of (quantized analog) signals (having integer
amplitudes) corresponding to the polyadic powers $\ell=1,2,3$, as follows
(using (\ref{af}) and (\ref{bl}))%
\begin{align}
\Psi_{tot}^{\left(  1\right)  }  &  =55\cdot f^{\left(  1\right)  }\left(
t\right)  ,\label{f1}\\
\Psi_{tot}^{\left(  2\right)  }  &  =171\cdot f^{\left(  2\right)  }\left(
t\right)  ,\label{f2}\\
\Psi_{tot}^{\left(  3\right)  }  &  =351\cdot f^{\left(  3\right)  }\left(
t\right)  , \label{f3}%
\end{align}
where $f^{\left(  1,2,3\right)  }\left(  t\right)  $ are different (or the
same) normalized signals. The recepient receives three (quantized analog) signals
(\ref{f1})--(\ref{f3}), and because he knows the normalized signals, he
immediately obtains the integer values $B_{1}=55$, $B_{2}=171$, $B_{3}=351$.
Inserting them into the system of quadratic equations (known to him ahead)
(\ref{ab1})--(\ref{ab3}), he (directly or using (\ref{mb})--(\ref{bb}))
derives the values $m=5$, $a=3$, $b=4$ and the desired element $y=m=5$ from
the initial plaintext (\ref{t}). The same procedure should be provided for
each element $y_{j}=y$ of the plaintext (\ref{t}), which completes its decryption.

\section{Conclusions}

Thus, we proposed a principally new encryption/decryption procedure based on
exploiting the signal processing. The main idea is to consider the signal
parameters as polyadic integers being representative of the fixed polyadic
ring, which is treated as some kind of polyadic discretization depending on
the ring integer characteristics. They allow us to transfer numerical
information from sender to recepient by submitting special sets of signals
with genuine properties agreed before. The recepient knows the rules and
equations to solve, to decrypt the initial plaintext. Security is achieved by using a mathematical process where public information is exchanged openly, but this information is useless without a corresponding piece of private, secret information that is never transmitted.